\documentclass[apj]{emulateapj}

\def\simgt{\lower.5ex\hbox{$\; \buildrel > \over \sim \;$}}
\def\simlt{\lower.5ex\hbox{$\; \buildrel < \over \sim \;$}}

\def\etal{{et~al.}}
\def\amin{\ifmmode^{\prime}\else$^{\prime}$\fi}
\def\asec{\ifmmode^{\prime\prime}\else$^{\prime\prime}$\fi}

\def\simgt{\lower.5ex\hbox{$\; \buildrel > \over \sim \;$}}
\def\simlt{\lower.5ex\hbox{$\; \buildrel < \over \sim \;$}}

\newcommand\chandra{{\it Chandra}}

\newcommand\xmm{{\it XMM-Newton}}
\newcommand\hess{{HESS}}
\newcommand\integral{{\it INTEGRAL\/}}
\newcommand\fermi{{\it Fermi\/}}

\newcommand\snr{G12.82$-$0.02}
\newcommand\tev{HESS~J1813$-$178}
\newcommand\igr{IGR~J18135$-$1751}
\newcommand\mev{0FGL~J1814.3$-$1739}
\newcommand\fgl{2FGL~J1814.1$-$1735c}

\newcommand\psr{PSR~J1813$-$1749}
\newcommand\asrc{AX~J1813$-$178}
\newcommand\cl{Cl~1813$-$178}

\slugcomment{}

\begin{document}

\title{Spin-Down Measurement of \psr: The Energetic Pulsar Powering \tev } 

\author{J.~P. Halpern, E.~V. Gotthelf, and F. Camilo}

\affil{Columbia Astrophysics Laboratory, Columbia University, 550 West
120th Street, New York, NY 10027-6601, USA}

\begin{abstract}

Two new X-ray timing observations of the 44.7~ms pulsar in \snr/\tev\ were
obtained with \chandra\ and \xmm\ to determine its precise spin-down rate.
With a period derivative of $\dot P = 1.265\times10^{-13}$, \psr\ is the
third most energetic pulsar in the Galaxy, having spin-down luminosity
$\dot E = 5.6\times10^{37}$~erg~s$^{-1}$.   Lack of pulsed detection in
a deep radio search from the Green Bank Telescope,
and in $\gamma$-rays from \fermi,
are reported.  We reconsider the distance to \psr/\snr\ in view of its
large X-ray measured column density, $N_{\rm H} = 10 \times10^{22}$~cm$^{-2}$,
which exceeds the visual extinction $A_V = 9.1$ to a young stellar cluster
at $d=4.8$~kpc that has been associated with it.  Although the distance
may well be larger, existing data do not constrain it further.
The small radiative output of \psr/\snr\ in all bands would not
exceed its spin-down power at any distance in the Galactic disk.
\end{abstract}

\section{Introduction}

\tev\ is one of the brightest and most compact TeV sources
discovered in the \hess\ Galactic Plane Survey \citep{aha05,aha06}.
It coincides with the young shell-type radio supernova remnant
(SNR) \snr\ and the 2--10~keV X-ray source \asrc\ \citep{bro05}.
It is also detected in 20--100~keV X-rays as the
\integral\ source \igr\ \citep{ube05}. High-resolution X-ray
studies resolved the X-ray emission into a point source and bright
surrounding nebula  \citep{fun07,hel07},
whose properties indicate that a young, energetic
rotation-powered pulsar is responsible for the
extended X-rays, and probably the TeV radiation as well.
A nearby young stellar cluster \cl\ at a kinematic
distance of 4.8~kpc was discovered by \citet{mes08,mes11},
who suggested this as a possible birth place of the pulsar progenitor.

We discovered 44.7~ms X-ray pulsations
from the point source in an \xmm\ observation \citep{got09}.
A $2.5\sigma$ detection of spin-down was also indicated 
in the long (98~ks) observation, suggesting that \psr\
is one of the most energetic pulsars.  In order
to confirm and refine the high spin-down rate, we obtained
two new X-ray observations, the results of which we report
in this Letter.

\section{X-ray Pulsar Observations}

A summary of the observations and results used here is
given in Table~\ref{data}. After the initial pulsar
discovery on 2009 March 27, \psr\ was observed a second
time by \xmm\ using the European Photon Imaging Camera (EPIC; \citealt{tur03}).
The EPIC pn CCD was operated in small-window mode ($4\farcm3 \times
4\farcm3$ field of view; 29\% dead time). This mode provides 5.7~ms
time resolution. Data were also
acquired with the two EPIC~MOS CCD cameras (MOS1 and MOS2). These 
were operated in full frame mode with a time resolution of
2.6~s, insufficient for pulsar timing, and are not used here.
In order to optimize the signal and minimize contamination
from the pulsar wind nebula (PWN), photons were extracted
in the 2--10~keV band from a radius of $20^{\prime\prime}$
around the pulsar position.

An observation was also obtained using the \chandra\
Advanced Camera for Imaging and Spectroscopy (ACIS)
operated in continuous-clocking (CC) mode to provide a time resolution
of 2.85~ms.   To achieve the fast timing in CC mode, one spatial
dimension of the CCD image (the row number) is lost.  To
minimize contamination from the PWN, five columns were extracted
around the pulsar position.  Although background was estimated
from nearby regions in both the \xmm\ and \chandra\ images,
it is difficult to subtract the background accurately
because of the structured PWN surrounding the pulsar.

\begin{figure}
\centerline{
\hfill
\includegraphics[height=1.\linewidth,angle=0,clip=true]{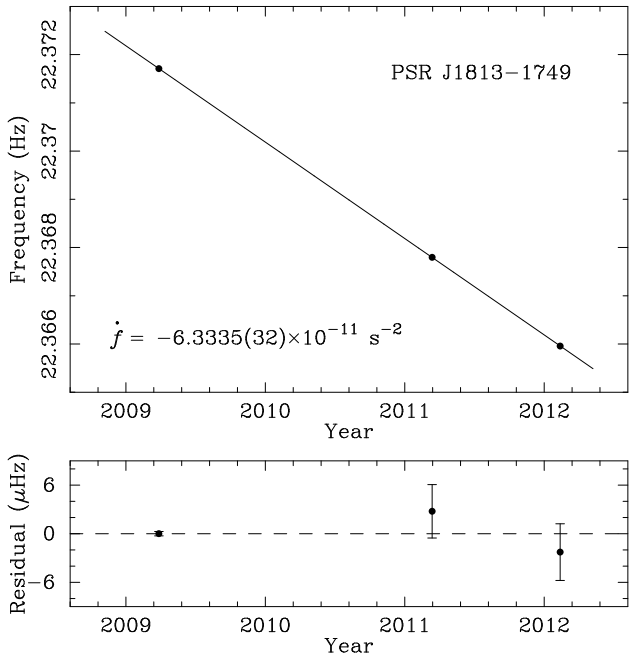}
\hfill
}
\caption{$\chi^2$ fit to the measured frequencies of \psr.
The bottom panel shows the residuals from the fit.
}
\label{fit}
\end{figure}

The photons arrival times were transformed to the solar system barycenter
in Barycentric Dynamical Time (TDB) using the \chandra\ measured coordinates
given in \citet{hel07} and listed in Table~\ref{timing}.  The best fitted
period was determined for each data set using the $Z^2_1$ (Rayleigh) test
\citep{str80,buc83}.  After determining the period derivative
accurately over the 3 yr span of the observations, the individual
observations were analyzed again using $Z^2_1$, this time with $\dot P$
held as a fixed parameter.  The best periods so determined are shown in
Figure~\ref{fit} and Table~\ref{data}, and the final
$\dot P = 1.26545(64) \times 10^{-13}$ is given in Table~2.
This is $<1\sigma$ from the originally estimated value in \citet{got09}.

\begin{figure}
\centerline{
\hfill
\includegraphics[height=1.25\linewidth,angle=0,clip=true]{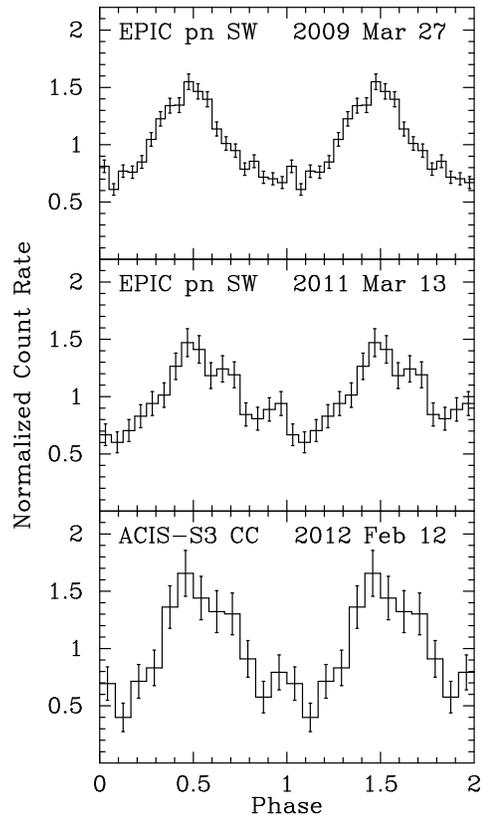}
\hfill
}
\caption{Pulse profiles of \psr\ in the 2--10~keV band,
folded according to the timing
parameters in Tables~\ref{data} and \ref{timing}.
Background from nearby regions of the PWN has been subtracted,
although it is not possible to do this exactly.
The pulsed fraction is $\approx 50\%$.
The pulses have been aligned arbitrarily in phase.}
\label{pulse}
\end{figure}

Figure~\ref{pulse} shows the individual folded light curves of
\psr\ from the three observations, with background subtracted
and the counts per bin normalized to~1.  Because the incoherent
timing solution fits only for frequency, not phase,
the pulse profiles were aligned arbitrarily.
The pulsed fraction is $\approx 50\%$.
The \chandra\
light curve provides a better estimate of the pulsed
fraction because it is possible to extract background
closer to the pulsar than it is in the \xmm\ image.

\begin{deluxetable*}{lcccccl}
\tablewidth{0pt}
\tabletypesize{\footnotesize}
\tablecaption{Timing Observations of  \psr }
\tablehead{
\colhead{Mission} & \colhead{Instrument/Mode} & \colhead{ObsID} & \colhead{Date (UT)} &
\colhead{Mid-Date (MJD)} & \colhead{Exp. (s)} & \colhead {Frequency (Hz)}
}
\startdata
\xmm\     & EPIC pn SW    & 0552790101 & 2009 Mar 27--28 & 54,918.14 & 98,360 & 22.37171236(27) \\
\xmm\     & EPIC pn SW    & 0650310101 & 2011 Mar 13    & 55,633.86 & 21,849 & 22.3677988(33) \\
\chandra\ & ACIS-S3 CC    & 12549      & 2012 Feb 12    & 55,969.24 & 20,077 & 22.3659583(35)
\enddata
\label{data}
\end{deluxetable*}

The derived physical parameters in the magnetic dipole model are spin-down power
$\dot E \equiv 4\pi^2 I \dot P P^{-3} = 5.6 \times 10^{37}$~erg~s$^{-1}$,
surface dipole magnetic field strength $B_s \equiv 3.2 \times 10^{19}\,
(P \dot P)^{1/2} = 2.4  \times 10^{12}$~G, and characteristic
age $\tau_c \equiv P/2\dot P = 5600$~yr.

\section{Search for Gamma-ray Pulsations}

Although a possible association of the \fermi\
source \mev\ with \tev\ was noted by \citet{abd09}, the
2FGL catalog \citep{nol12} indicates only a confused source
\fgl\ with a 95\% error radius of $4.\!^{\prime}8$ that lies
$16.\!^{\prime}6$ from \tev.  \citet{mar11} derived a flux
upper limit of $<3\times10^{-11}$ erg~cm$^{-2}$~s$^{-1}$
($> 100$~MeV) from \fermi\ data at the position of \psr.
The \fermi\ luminosity limit (assumed isotropic) in Table~\ref{flux}
is not inconsistent with the trend and scatter among efficiencies
of $\gamma$-ray pulsars as seen in Figure~6 of \citet{abd10a}
or Figure~2 of \citet{mar11},
although it is below the best fit of the high $\dot E$ population.
The \fermi\ upper limit lies above an extrapolation of the
\hess\ spectrum, so the \fermi\ data are not sensitive
enough to constrain the $\gamma$-ray emission mechanism.

For completeness, we have searched the \fermi\ data at the position
of \psr\ for pulsations using the X-ray measured timing parameters
and their errors in Table~\ref{timing} to restrict the parameters
of the search.  To match the time span of the X-ray observations,
we extracted LAT data from 2009~March~27 to 2012~February~12. 
These data were reprocessed using the
``Pass~7'' event reconstruction algorithm. We selected only class~2
``diffuse'' photons for zenith angles $<100^{\circ}$ and restricted the
energy range to $>500$~MeV to minimize solar limb and diffuse
Galactic $\gamma$-ray background contamination. The photon arrival
times were corrected to the solar system barycenter.
For our nominal pulsar search we extracted
photons from a $1.\!^{\circ}5$ radius aperture.

Because the sparse X-ray observations do not yield a phase-connected
ephemeris, unknown timing noise could limit the time span of a
coherent pulsar search.  Accordingly,
we grouped the photons into intervals of 10 days and performed
a $Z^2_5$ test (summed over five harmonics) to allow for a
complex profile with narrow features. For each time interval, we
searched a range of $f,\dot f$ centered on interpolated values at
the test epoch using the timing parameters in Table~\ref{timing}.
The search range corresponds to three times the $1\sigma$
uncertainty in each of $f$ and $\dot f$.  We oversampled in each
of these parameters by a factor of three in the highest test harmonic
to ensure sensitivity to sharp pulse profiles.

None of the 106 resulting searches yielded a significant
signal above the noise.  We also summed the power from these searches
for each $f,\dot f$ pair, increasing the signal sensitivity by a
factor of $\sim 10$.  No significant signal was apparent.
To explore a range of instrumental parameters that might
be more sensitive, we repeated our search for extraction radii
$1.\!^{\circ}0,1.\!^{\circ}5,2.\!^{\circ}0,2.\!^{\circ}5$,
low-energy cutoff $E_{\rm min}=100,300,500$ MeV,
$Z^2_n$ harmonics $n=1,3,5$, and time
intervals 10,30,60,120 days. We also used the $H$-test \citep{dej89};
from a range of harmonics, this test selects the $n$ that results in
the most significant $Z^2_n$ value. Despite the expanded range of
parameter space, no signal stronger than the expected noise is detected 
in each search.

\begin{deluxetable}{ll}
\tablewidth{0pt}
\tablecaption{\label{ephem}Timing Parameters for \psr }
\tablehead{
\colhead{Parameter}   &
\colhead{Value}   }
\startdata                                       
R.A. (J2000)\tablenotemark{a}              & $18^{\rm h}13^{\rm m}35.\!^{\rm s}166$\\
Decl. (J2000)\tablenotemark{a}             & $-17\arcdeg49'57.\!^{\prime\prime}48$\\
Epoch (MJD)                                & 54918.14                        \\
Period, $P$ (s)                            & 0.04469930526(54)               \\
Period derivative, $\dot P$                & $1.26545(64)\times10^{-13}$      \\
Frequency, $f$ (Hz)                        & 22.37171236(27)                 \\
Frequency derivative, $\dot f$ (Hz s$^{-1}$)& $-6.3335(32)\times10^{-11}$      \\
Characteristic age, $\tau_c$ (yr)          & 5600                            \\
Spin-down luminosity, $\dot E$ (erg\,s$^{-1}$) & $5.6\times10^{37}$       \\
Surface dipole magnetic field, $B_s$ (G)       & $2.4\times10^{12}$
\enddata
\tablecomments{\footnotesize $1\sigma$ uncertainties are given.}
\tablenotetext{a}{\footnotesize \chandra\ ACIS-I position from \cite{hel07}.}
\label{timing}
\end{deluxetable}

\begin{deluxetable*}{lcccccl}
\tablewidth{0pt}
\tabletypesize{\footnotesize}
\tablecaption{Luminosity and Radiative Efficiency of \psr/\snr }
\tablecolumns{7}
\tablehead{
\colhead{Mission} & \colhead{Energy Band} & \colhead{Source}
& \colhead{Flux\tablenotemark{a}} & \colhead{Luminosity\tablenotemark{b}}
& \colhead{Efficiency} & \colhead{Flux Reference} \\
\colhead{} & \colhead{} & \colhead{} &  \colhead{(erg~cm$^{-2}$~s$^{-1}$)}
& \colhead{(erg~s$^{-1}$)} & \colhead{} & \colhead{}
}
\startdata
\chandra\               & 2--10~keV  & \asrc & $7.2 \times 10^{-12}$
 & $2.0 \times 10^{34} $ & $3.5\times10^{-4}$ & \cite{hel07} \\
\integral\              & 20--100~keV & \igr\ & $2.6 \times 10^{-11}$
 & $7.0 \times 10^{34}$  & $1.3\times10^{-3}$ & \cite{bir10} \\
\fermi\                 &  $>100$~MeV & . . . & $<3 \times 10^{-11}$ 
& $<8.3 \times 10^{34}$  & $<1.5\times10^{-3}$ & \cite{mar11} \\
\hess\                  &  $>200$~GeV & \tev\ & $1.8 \times 10^{-11}$
& $4.9 \times 10^{34}$   & $8.8\times10^{-4}$ & \cite{aha06}
\enddata
\tablenotetext{a}{Unabsorbed flux, including pulsar and PWN.}
\tablenotetext{b}{\footnotesize Luminosity assuming $d=4.8$~kpc.}
\label{flux}
\end{deluxetable*}

\section{Radio Pulsar Search}

The location of \psr\ was searched unsuccessfully for radio
pulsations at the Parkes telescope in 2005, with a resulting limit
on period-averaged flux density at 1.4\,GHz of $S_{1.4}<0.07$\,mJy,
assuming a 10\% duty cycle \citep{hel07}.  In 2009 we did two,
more sensitive observations at the NRAO Green Bank Telescope (GBT).

On 2009 May 25 we used the GUPPI
spectrometer\footnote{https://safe.nrao.edu/wiki/bin/view/CICADA/GUPPiUsersGuide} to sample an 800~MHz band centered at 2~GHz, recording data from each
of 512 frequency channels every 0.163~ms.  The integration lasted for
1.8~hr.  On 2009 August 17 we did a similar observation at the GBT,
but with an integration time of 8.5~hr.  We analyzed both data sets
using standard pulsar search techniques implemented in PRESTO \citep{ran01},
but found no pulsar candidates.  The data were searched in
dispersion measure up to 3360~pc~cm$^{-3}$, twice the total Galactic
value predicted for this line of sight by the \citet{cor02}
electron distribution model.

The flux density limits for these observations at 2 GHz and for
the period of the pulsar, assuming a 10\% pulsar duty cycle, are
0.013~mJy and 0.006~mJy, respectively. Converting to 1.4~GHz,
assuming a typical pulsar spectral index of $-1.6$, these correspond
to $S_{1.4}<0.023$~mJy and $S_{1.4}<0.01$~mJy, respectively.  For an
assumed distance of 4.8~kpc (see below), the luminosity
limit of our most sensitive radio search is $L_{1.4} \equiv S_{1.4}
d^2 < 0.2$ mJy~kpc$^2$.  With one exception \citep{abd10b}, all
young pulsars with detected radio pulsations are more luminous than this
\citep[see][]{cam09a,cam09b}, which holds true even for a distance
of 7~kpc.

A curious discovery of a time-variable point source
in VLA 4.85~GHz data at the location of \psr\ 
was reported by \citet{dzi10}.  The measured flux
density of $0.18 \pm 0.02$~mJy in 2006 February
was $1.9 \pm 0.7$ times larger than an upper limit
obtained in 2009 March.  We are not aware if the
single detection has been duplicated. 
At 4.8~GHz, a typical rotation-powered radio pulsar is some seven times
fainter than at 1.4~GHz.  Our three flux density limits extrapolated
to 4.8~GHz are thus a factor of approximately 20--120 times smaller
than the time-variable compact source detection claimed by \citet{dzi10},
which therefore almost certainly does not arise from pulsed
emission\footnote{Radio magnetars display both flat spectra and highly
variable emission \citep[e.g.,][]{cam07}, but \psr\ is
not a magnetar.}.

\section{Discussion}

\subsection{Distance, Environment, and Associations}

The total Galactic \ion{H}{1} column in the direction of \snr\ is
only $N_{\rm H} = 1.9 \times 10^{22}$~cm$^{-2}$ \citep{dic90} or
$1.3 \times 10^{22}$~cm$^{-2}$ \citep{kal05}, while \citet{bro05}
find that the molecular column density between 0 and 4~kpc
is $N({\rm H}_2) = 8 \times 10^{22}$~cm$^{-2}$ by integrating
the CO data of \citet{dam01}.  Higher resolution CO mapping
by \citet{fun07} using NANTEN, with a beam size of $2.\!^{\prime}6$
and a grid spacing of $4^{\prime}$,
found $N({\rm H}_2) = 9 \times 10^{22}$~cm$^{-2}$ at the position
of \snr. Thus, the total X-ray measured column density to \snr\ of
$N_{\rm H} = (10\pm 1)\times10^{22}$~cm$^{-2}$
\citep{fun07,hel07} can be accounted for if \snr\ lies at
a distance $\ge 4$~kpc.  From this, and the apparent
free--free absorption of \snr\ by the W33 star-forming region
at a distance of 4.3~kpc, previous authors
\citep{bro05,hel07,got09} have concluded that
the distance to \snr\ is $\approx 4.5$~kpc.

\citet{mes08,mes11} discovered a young stellar cluster
in Two Micron All Sky Survey
infrared images on the edge of the W33 complex near \snr.
Dubbed \cl, the cluster is centered $4.\!^{\prime}4$ southwest of \snr,
most of its members falling within $3.\!^{\prime}5$ of the cluster center.
\citet{mes11} concluded that it is one of several
clusters belonging to the W33 complex.  They derived a
spectrophotometric distance to \cl\ of $3.6 \pm 0.7$~kpc,
and a kinematic distance of $4.8\pm 0.3$~kpc from the
radial velocity ($v_{\rm LSR}=+62\pm 4$ km~s$^{-1}$)
of the brightest star in the cluster.
Rich in massive, young stars, and containing a second
SNR G12.72$-$0.00, this seems an ideal
birth place for \psr.
\citet{mes11} determined an age of 4--4.5~Myr for the cluster,
with a spread in age of 1~Myr based on the simultaneous presence
of red supergiants, Wolf--Rayet, and Of stars.
The total mass of the cluster is $\ge 10,000\ M_{\odot}$.  
It is plausible that the progenitors of both \snr\ and G12.72--0.00
were born in the cluster, and had masses similar to the
estimated main-sequence turnoff, $25-35 M_{\odot}$.
Only 13 such massive clusters are known in the Galaxy \citep{mes09}.

However, evidence that may contradict an association of \snr\ with
\cl\ are the discrepant absorption column densities to the two objects.
The average visual extinction to \cl\ of $A_V = 9.1$ \citep{mes11},
estimated from the infrared colors of 25 of its member stars,
corresponds to an equivalent X-ray absorption of only
$N_{\rm H} = 1.6 \times 10^{22}$~cm$^{-2}$ according to the relation
of \citet{pre95}.  The highest extinction for a cluster
member is $A_V = 17$.  Either of these extinction 
values do not account for the X-ray measured
$N_{\rm H} = (10\pm 1) \times 10^{22}$~cm$^{-2}$ to \snr.
Although the difference may be made up by half of the molecular
column of $N({\rm H}_2) = 9 \times 10^{22}$~cm$^{-2}$
measured by \citet{fun07}, this would require
nearly all of the X-ray measured column density to \snr\
to lie behind \cl.   This means that \snr\
is not constrained to be at the kinematic distance of \cl.
There is no evidence that the SNR is interacting 
physically with molecular gas that could be associated with it.
In addition, it is likely that some of the molecular hydrogen
on the line of sight, which has $v_{\rm LSR} \le +60$~km~s$^{-1}$,
is in the far branch of the double-valued radial velocity curve
at $12-16$~kpc.  This means that \snr\ could be anywhere
from $5-12$~kpc distant, unconstrained by the distribution 
of molecular gas.   If so, it could be expected to have
\ion{H}{1} absorption features up to $v_{\rm LSR} \approx +170$~km~s$^{-1}$,
the tangent point velocity at $d = 8.3$~kpc.
\citet{bro05} report that
no \ion{H}{1} absorption features are seen toward \snr\
at $v_{\rm LSR} \ge +55$~km~s$^{-1}$ in data from
the Southern Galactic Plane Survey, although the signal
to noise is low; therefore, their absence is inconclusive.

Indirect evidence for a larger distance comes from
a comparison with the absorption to the bright
LMXB GX~13+1 that lies only $0.\!^{\circ}7$  from \snr\
along the Galactic plane.  The total Galactic \ion{H}{1} column
densities are substantially the same along these lines of sight.
Converting the CO intensity of \citet{dam01} to ${\rm H}_2$ column density, we
find $N({\rm H}_2) = 5 \times 10^{22}$~cm$^{-2}$ in the direction of GX~13+1. 
The distance to GX~13+1 was estimated as $7\pm 1$~kpc
\citep{ban99} from the IR spectroscopic classification (K5III)
and reddening ($A_V=13.2-17.6$) of its companion star,
while its X-ray column density measured by ASCA
is $N_{\rm H} = (2.9\pm 0.1) \times 10^{22}$ cm$^{-2}$ \citep{ued01}.
The X-ray measured absorption is consistent with being
due mostly to molecular material.
As this is less than one-third the X-ray measured $N_{\rm H}$ to \snr,
it suggests that the latter may be farther than 7~kpc.

Finally, we note that these comparisons involving individual
stars are subject to the caveat that extinction and CO may be
clumpy on small scales that are not resolved by the
CO surveys. Lacking certainty, we therefore
retain the previously adopted distance of 4.8~kpc
to calculate the luminosities in Table~\ref{flux}.
Even at a distance of 15~kpc, or a factor of 10 higher
in luminosity than assumed in Table~\ref{flux}, the radiative
efficiencies in all bands would be plausible for such
an energetic pulsar.  The \fermi\ upper limit would be
consistent with the GeV luminosities of the
bulk of the high $\dot E$ pulsar population.
The TeV luminosity would still
be $<1\%$ of $\dot E$, although \psr\ would then
have the highest luminosity TeV PWN in the Galaxy.  

\subsection{Age and Energetics}

HESS~J1813$-$178 is one of the more compact TeV sources
to be associated with an SNR.  The Gaussian extent of the source is only
$\sigma = 2.\!^{\prime}2 \pm 0.\!^{\prime}4$ \citep{aha06},
while the radio shell of \snr\ is
$2.\!^{\prime}5$ in diameter \citep{bro05},
corresponding to a radius of 1.7~pc at a distance of 4.8~kpc.
The small size can be explained by a young age. 
Although the characteristic age of \psr\ is 5600~yr,
its true age may be significantly smaller than this
if it were born with an initial period not much less
than its present period.  \cite{bro05} estimate a free-expansion
age for \snr\ of only $340\,(d/4.8~{\rm kpc})\,(v_s/5000~{\rm km~s}^{-1})$~yr,
which would be consistent with having swept up little
interstellar matter at the assumed distance.
Even if at $d=20$~kpc, they note that the Sedov--Taylor 
age for an explosion energy of $E_0 = 10^{51}$~erg and an
ambient density of 1~cm$^{-3}$ is $\approx 2500$~yr.
Using the pulsar plus PWN fluxes, the
ratio $L_x(2-10\ {\rm keV})/L_{\gamma}\ (>200\ {\rm GeV})$
is $\approx 0.4$, which is typical
for the $\dot E$ and characteristic age of \psr\ \citep{mat09}.

A detailed model of the evolution and multiwavelegth
emission from \tev\ was constructed by \citet{fan10},
including particle acceleration in both the PWN and SNR.
They assumed $d=4.7$~kpc, an underluminous explosion
$E_0 = 0.5 \times 10^{50}$~erg,
an ambient density of 1.4~cm$^{-3}$, and a present age
of 1200~yr.  At this stage the reverse shock
has not yet reached the PWN, which has a radius of 0.7~pc.
The results are that the PWN is responsible
for the X-ray and the TeV emission, while the SNR
shell is only detectable in radio.  The TeV emission
is inverse Compton scattered microwave background
or IR photons from dust,
while the stellar light is a small contributor.
Energy densities for IR and starlight were
taken to be $U_{\rm IR} = 1.0$~eV~cm$^{-3}$
and $U_* = 1.5$~eV~cm$^{-3}$, respectively. 

A justification for these energy densities was not given,
but an assumed association with \cl\ is not
consistent with such small
values.  The massive stars in the cluster are
very luminous; the total of just the 25 young cluster
members is $L_* = 1.04 \times 10^7\ L_{\odot}$ at
the kinematic distance of 4.8~kpc \citep{mes11}.
If \snr\ is at the projected distance of $4.\!^{\prime}4$
($r = 6.2$~pc) from the cluster center, the energy density in its vicinity
can be approximated as $U_* = L_*/4\pi r^2 c = 180$
eV~cm$^{-3}$.  Even if this is reduced by a factor $\sim 2$
due to projection effects, it is still much larger than
$U_{\rm IR} + U_* = 2.5$ eV~cm$^{-3}$ assumed by \citet{fan10}.
This suggests that either their model should be revised
to account for the energy density in starlight from \cl,
or more likely, \snr/\tev\ lies at a larger distance than \cl.
If the distance is much larger, this would also allow a
more energetic supernova explosion.

\section{Conclusions}

We obtained new X-ray timing observations of \psr, which
precisely refine the previously suggested high spin-down rate.
With $\dot E = 5.6 \times 10^{37}$ erg~s$^{-1}$, \psr\
is the third most energetic pulsar in the Galaxy
after the Crab and the recently discovered
PSR~J2022+3842 \citep{arz11}.  Uncertainty about the distance
to \psr/\snr\ arises from discrepant absorption column densities
to an assumed host stellar cluster and another bright X-ray
source along the line of sight.  We conclude that the distance
may be larger than the previously adopted 4.8~kpc, but it is
difficult to quantify further.  The small radiative efficiency
of \psr/\snr\ in all bands allows any distance in the Galactic
disk without exceeding its spin-down power.  A more sensitive observation
of 21~cm \ion{H}{1} absorption against \snr\ is needed 
to pin down its distance.
It would be worthwhile to construct models for the SNR evolution of
\snr\ and the TeV emission from \tev\ for a larger distance.

\acknowledgements

We thank Roland Kothes for helpful discussions.
This investigation is based on observations obtained with \xmm, an ESA
science mission with instruments and contributions directly funded by
ESA Member States, and NASA.  Support for this work was provided by
NASA through \chandra\ award SAO G01-12075X issued by the Chandra X-ray
Observatory Center, which is operated by the Smithsonian Astrophysical
Observatory for and on behalf of NASA under contract NAS8-03060, and
by \fermi\ guest investigator grant NNX11AO36G.

\end{document}